\documentstyle[12pt]{article}
\def\be{\begin{equation}}
\def\ee{\end{equation}}
\def\bea{\begin{eqnarray}}
\def\eea{\end{eqnarray}}

\begin{document}

\begin{center}
{\Large{\bf Ghosts in the Matter Forms}}

\vskip .5cm
{\large Davoud Kamani}
\vskip .1cm
 {\it Institute for Studies in Theoretical Physics and
Mathematics (IPM)
\\  P.O.Box: 19395-5531, Tehran, Iran}\\
{\sl e-mail: kamani@theory.ipm.ac.ir}
\\
\end{center}

\begin{abstract}

In this paper we study the matter form of the conformal and
super-conformal ghosts action. 
That is, the ghost fields will be expressed
in terms of some scalar and spinor fields. 
Thus, we obtain a two-dimensional
covariant action in the matter form, $i.e.$ $S_g$.
The Poincar\'e-like symmetries and
various supersymmetries of this covariant action are analyzed.
The signatures 10+2 and 11+3 for the total
target space of the superstring theory also will be discussed.

\end{abstract}



\vskip .5cm
\newpage
\section{Introduction}
In the recent years, string theories can be understood by assuming
the existence of higher dimensional target spaces
\cite{1,2,3,4,5,6}.
The eleven-dimensional M-theory \cite{3} and the twelve-dimensional
theory known as F-theory \cite{4}, are examples of this context.
The analysis of the super $p$-brane scanning allows spacetimes with
non-Lorentzian signatures \cite{5}.
In other words, there are several models which have more than one
``time'' coordinate \cite{2,4,5,6}.
In particular, twelve-dimensional theories and their invariances
with respect to the $SO(10,2)$ rotations have been investigated \cite{2}.

Since the superstring
possesses gauge symmetries, namely worldsheet reparametrization invariance,
the procedure for the path integral quantization of the
superstring is the Faddeev-Popov method. On the other hand, ghosts
are quantum fields used to give a functional integral representation
of the Faddeev-Popov determinant \cite{7}. They also have an 
important role in the BRST quantization \cite{8}.
If two or more-time world is real, we should be able to formulate the
superstring theory in the language of two or more-time physics
without the conformal and super-conformal ghosts.

We shall express the action of the super-conformal and conformal ghosts
in the covariant form of the matter fields.
Therefore, the ghost fields have expressions in terms of the bosonic
and fermionic fields. Quantum consistency of the ghosts action
in the matter form and initial form will be shown.
The matter form of the action enables us to study
the symmetries of the theory. 
Two of these symmetries are $N=1$ and $N=2$
supersymmetries. However, in
these formulations the superstring lives in the 
11+3 or 10+2 dimensional spacetimes without any ghost field.

Besides the $N=2$ supersymmetry, the theory is invariant
with respect to two Poincar\'e-like symmetries and two other 
supersymmetries. For each of these symmetries there are two
conserved currents. That is, each symmetry is described by
the product of two distinct groups.

This paper is organized as follows. In section 2, by introducing
some vectors, the super-conformal and conformal ghosts
and their action will be expressed in terms of the matter fields.
In section 3, the superstring action beyond the dimension ten
will be presented. In section 4,
Poincar\'e-like symmetries, bi-supersymmetries and $N=2$ supersymmetry
of the new form of the ghosts action will be studied. In section 5, the
signature 11+3 for the total target spacetime 
of the superstring theory will be discussed.
\section{Matter form of the ghosts action}
The superstring with the worldsheet supersymmetry has the action 
\bea
S=-\frac{1}{4\pi \alpha'}\int d^2\sigma \bigg{(}
G_{\mu\nu}(\eta^{ab} \partial_a X^\mu \partial_b X^\nu
-i{\bar \psi}^\mu \rho^a \partial_a \psi^\nu )\bigg{)}+S_g\;,
\eea
where $S_g$ is sum of the conformal and super-conformal
ghosts actions, $i.e.$ $S_g = S_{scg}+S_{cg}$.
As we know, the spacetime corresponding to the action (1) 
has the dimension ten with the signature 9+1.
Now we proceed to study the covariant matter form of the action $S_g$.
\subsection{The super-conformal ghosts}
The super-conformal ghosts have the action 
\bea
S_{scg} = \frac{1}{2\pi \alpha'}\int d^2\sigma (\beta \partial_+\gamma
+{\tilde \beta}\partial_-{\tilde \gamma})\;,
\eea
where $\partial_{\pm}=\frac{1}{2}(\partial_\tau \pm \partial_\sigma)$.
In this action the fields $\beta$ and $\gamma$
(and also ${\tilde \beta}$ and
${\tilde \gamma}$) enter symmetrically, despite the asymmetrical
appearance of them. This is due to the flat worldsheet.

Consider the quantities
$\{Y^P, {\tilde Y}^P, \partial_+Z^P, \partial_-{\tilde Z}^P\}$.
We express them in terms of the variables
$\{\partial_-X^p,
\partial_+X^p, \partial_-{\tilde X}^p,\partial_+{\tilde X}^p\}$
as in the following
\bea
&~&Y^P =e^P_{\;\;\;p} \partial_-X^p\;,
\nonumber\\
&~&\partial_+Z^P =e^P_{\;\;\;p} \partial_+X^p\;,
\nonumber\\
&~&{\tilde Y}^P = e^P_{\;\;\;p} \partial_+{\tilde X}^p\;,
\nonumber\\
&~&\partial_-{\tilde Z}^P = e^P_{\;\;\;p} \partial_-{\tilde X}^p\;,
\eea
where the set $\{e^P_{\;\;\;p}\}$ denotes 
the two-dimensional vielbeins with
$p,P \in \{1,2\}$. The other vielbeins (e.g., those that connect
$Y^P$ to $\partial_+X^p$ and $\partial_{\pm}{\tilde X}^p$) are zero.
The equations (3) imply that only four coordinates of the set
$\{Y^P , {\tilde Y}^P,Z^P , {\tilde Z}^P\}$ are independent. For example,
for the constant vielbeins there are $\partial_+Y^P=\partial_+\partial_-Z^P$
and $\partial_-{\tilde Y}^P=\partial_+\partial_-{\tilde Z}^P$.

We demand the action (2) to be
\bea
S_{scg}= \frac{1}{2\pi \alpha'}\int d^2\sigma \bigg{(}\eta_{PQ}
(Y^P \partial_+Z^Q + {\tilde Y}^P \partial_-{\tilde Z}^Q)
\bigg{)}\;,
\eea
where the metric $\eta_{PQ}$ has the Lorentzian signature, $i.e.$
$\eta_{PQ}={\rm diag}(-1,1)$. This action contains the bosonic fields
and has the feature of the action (2). Therefore, the fields of (4)
have the roles of the super-conformal ghosts. 
Expanding the action (4) gives two copies of the action (2). 
On the other hand, the action (4) 
has all symmetries of the action (2).

Define the two-dimensional metric $G_{pq}$ as
\bea
&~&G_{pq}=\sum_{P,Q}(\eta_{PQ} e^P_{\;\;\;p} e^Q_{\;\;\;q})\;,
\nonumber\\
&~&\eta^{PQ}=\sum_{p,q}(G^{pq}e^P_{\;\;\;p} e^Q_{\;\;\;q})\;.
\eea
Therefore, the action (4) takes the form
\bea
S_{scg}=-\frac{1}{8\pi \alpha'}\int d^2\sigma \bigg{(}
G_{pq}\eta^{ab}( \partial_a X^p \partial_b X^q
+\partial_a {\tilde X}^p \partial_b {\tilde X}^q)\bigg{)}\;.
\eea
This action corresponds to the manifold ${\cal{M}}_g$ with the
coordinates $\{X^p , {\tilde X}^p\}$ and the metric
\bea
{\bar G}= \left( \begin{array}{cc}
\frac{1}{2}G_{pq} & 0\\
0 & \frac{1}{2}G_{pq}
\end{array} \right).
\eea
We call ${\cal{M}}_g$ as the ghosts manifold. 
The metric $G_{pq}$ provides
a background for the string which propagates in this manifold.
This form of the action is covariant with respect to the worldsheet
indices $\{a,b\}$ and the manifold indices $\{p,q\}$.
We shall see that there are some conditions on the fields of
the action (6) which
give the equality of the degrees of freedom of this action with the
action (2).

In a system with the $D$-dimensional Poincar\'e symmetry $ISO(D-1 , 1)$,
the conformal symmetry is $SO(D,2)$. In fact, when one considers the
conformal symmetry, the symmetry $SO(1,1)$ is added to the original
global symmetry. Thus, besides the usual time coordinate, the
conformal symmetry introduces another time coordinate. These imply that
the ghosts manifold is product of two identical
copies of a two-dimensional spacetime. In other words, we have a 
four-dimensional spacetime with the signature 2+2.
We shall see that the 1+1 interpretation for the signature of this
spacetime also is possible.

In fact, the action (4) is auxiliary. 
Since the actions (2) and (4) have the same feature 
we can write
\bea
&~&\beta \partial_+\gamma =\eta_{PQ} Y^P\partial_+ Z^Q \;,
\nonumber\\
&~&{\tilde \beta}\partial_-{\tilde \gamma}
=\eta_{PQ}{\tilde Y}^P \partial_-{\tilde Z}^Q\;.
\eea
Let $V^P$ and ${\tilde V}^P$ be two unit vectors, $i.e.$, 
\bea
\eta_{PQ}V^P V^Q=\eta_{PQ}{\tilde V}^P {\tilde V}^Q=1\;.
\eea
Insertion of these vectors in the left-hand sides of the equations (8)
leads to the equations
\bea
(\beta V)^T\eta (V\partial_+\gamma)=Y^T\eta\partial_+ Z\;,
\nonumber\\
({\tilde \beta} {\tilde V})^T\eta ({\tilde V}\partial_-
{\tilde \gamma})={\tilde Y}^T\eta\partial_- {\tilde Z}\;.
\eea
One solution of these equations is
\bea
&~&\beta V=Y\;,
\nonumber\\
&~& V\partial_+ \gamma=\partial_+ Z +U\;,
\nonumber\\
&~&{\tilde \beta} {\tilde V}={\tilde Y}\;,
\nonumber\\
&~&{\tilde V}\partial_-{\tilde \gamma}=\partial_- {\tilde Z}
+{\tilde U}\;,
\eea
where the vectors $U^P$ and ${\tilde U}^P$ are 
perpendicular to $Y^P$ and ${\tilde Y}^P$, respectively,
\bea
\eta_{PQ}Y^P U^Q = \eta_{PQ}{\tilde Y}^P {\tilde U}^Q =0.
\eea
However, for the next purposes, we assume they are not
perpendicular to $V^P$ and ${\tilde V}^P$, $i.e.$,
\bea
\eta_{PQ}V^P U^Q \neq 0 \;\;\;,\;\;\; \eta_{PQ}{\tilde V}^P 
{\tilde U}^Q \neq 0.
\eea

Other solutions for the equations (10) are possible. For example,
they can be written in the form
$(V\partial_+\gamma)^T\eta(\beta V) =Y^T\eta\partial_+ Z$
and
$({\tilde V}\partial_-{\tilde \gamma})^T \eta ({\tilde \beta}{\tilde V})
={\tilde Y}^T\eta\partial_- {\tilde Z}$. 
The solution of these equations is different from (11). 
We consider only the solution (11).

In terms of the fields of the action (4) and also in terms of the
coordinates $\{X^p , {\tilde X}^p\}$ the super-conformal ghosts
have the following expressions
\bea
&~&\beta =\eta_{PQ} V^P Y^Q = v_p \partial_-X^p\;,
\nonumber\\
&~& \partial_+ \gamma=\eta_{PQ} V^P(\partial_+ Z^Q + U^Q) = v_p
\partial_+ X^p + v.u\;,
\nonumber\\
&~&{\tilde \beta}=\eta_{PQ} {\tilde V}^P{\tilde Y}^Q
={\tilde v}_p \partial_+ {\tilde X}^p\;,
\nonumber\\
&~&\partial_-{\tilde \gamma}=\eta_{PQ}
{\tilde V}^P(\partial_- {\tilde Z}^Q + {\tilde U}^Q)
={\tilde v}_p \partial_- {\tilde X}^p + {\tilde v}.{\tilde u}\;,
\eea
where the vectors $v_p$ and ${\tilde v}_p$ have the definitions
\bea
&~&v_p=\eta_{PQ}e^P_{\;\;\;p} V^Q\;,
\nonumber\\
&~&{\tilde v}_p=\eta_{PQ}e^P_{\;\;\;p} {\tilde V}^Q\;,
\eea
similarly for the vectors $u_p$ and ${\tilde u}_p$.
The inner product $v.u$ is defined by
$v.u = G_{pq}v^p u^q$. According to the equations (13),
$v.u$ and ${\tilde v}.{\tilde u}$ are nonzero. Note that 
$v_p$ and ${\tilde v}_p$ 
also are unit vectors, $i.e.$ $G^{pq}v_pv_q =
G^{pq}{\tilde v}_p{\tilde v}_q = 1$. The equations (14) imply that the
super-conformal ghosts can be seen as linear combinations of some
scalar fields.

To understand more about the mappings (14), 
let consider the bosonization of the 
super-ghosts $\beta$ and $\gamma$,
\bea
\beta=e^{-\phi}\partial_- \xi\;,\;\;\; \gamma=
e^{\phi}\eta\;,\;\;\; \xi=e^{\zeta}\;,\;\;\; \eta=e^{-\zeta}\;,
\eea
where the bosonized super-ghosts are 
$\phi$, $\xi$, $\eta$ and $\zeta$ \cite{9}.
A similar construction for the ghosts $b$ and $c$ 
was carried out in Ref.\cite{10}. 
As we see, $\beta$ depends on the 
derivative of a field, while $\gamma$ is 
independent of any derivative. 
This also is true for $\beta$ and $\gamma$ in the equations (14). 
For the field $\gamma$ also see the first equation 
of (31). Therefore, the equations (14) can be interpreted as 
a kind of bosonization. 
Thus, the fields $X^1$ and $X^2$ have the following relations with the 
bosonized super-ghosts $\phi$ and $\zeta$,
\bea
v_p \partial_- X^p = e^{-\phi+\zeta}\partial_-\zeta\;,
\nonumber\\
v.u + v_p \partial_+ X^p = \partial_+ e^{\phi-\zeta}\;.
\eea
Similar interpretation also holds for the left-moving fields.

According to the equations (11), it is possible to express the fields
$\{X^p , {\tilde X}^p\}$ in terms of the ghost fields
\bea
&~&\partial_- X^p = v^p\beta\;,
\nonumber\\
&~&\partial_+ X^p = v^p\partial_+\gamma - u^p\;,
\nonumber\\
&~&\partial_+ {\tilde X}^p = {\tilde v}^p{\tilde \beta}\;,
\nonumber\\
&~&\partial_- {\tilde X}^p = {\tilde v}^p\partial_-{\tilde \gamma}
-{\tilde u}^p\;.
\eea
The contravariant vectors are
$v^p=e^{\;\;\;p}_P V^P$, ${\tilde v}^p=e^{\;\;\;p}_P{\tilde V}^P$,
$u^p=e^{\;\;\;p}_P U^P$ and ${\tilde u}^p=e^{\;\;\;p}_P{\tilde U}^P$,
where the matrix $e^{\;\;\;p}_P$ is inverse of the vielbein matrix
$e^P_{\;\;\;p}$. These matrices satisfy the relations
$e^P_{\;\;\;p}e^{\;\;\;p}_{P'}=\delta^P_{\;\;\;P'}$
and $e^P_{\;\;\;p}e^{\;\;\;p'}_P=\delta^{\;\;\;p'}_p$.

The solutions of the equations of motion of the action (6) have the
general forms $X^p = X^p_R + X^p_L$ and ${\tilde X}^p =
{\tilde X}^p_R + {\tilde X}^p_L$. Half of the degrees of freedom of
$X^p$ and ${\tilde X}^p$ correspond to the super-conformal ghosts.
Now consider the unit vectors $V'$ and ${\tilde V'}$ which are
perpendicular to the vectors $V$ and ${\tilde V}$, respectively.
The inner products of the vectors $V'$ and 
${\tilde V'}$ with the vectors in the
equations (11) lead to the conditions
\bea
&~&\eta_{PQ} V'^P Y^Q = v'_p \partial_-X^p=0\;,
\nonumber\\
&~& \eta_{PQ} V'^P(\partial_+ Z^Q + U^Q) = v'_p \partial_+ X^p
+ v'.u=0\;,
\nonumber\\
&~&\eta_{PQ} {\tilde V}'^P {\tilde Y}^Q
={\tilde v'}_p \partial_+ {\tilde X}^p=0\;,
\nonumber\\
&~&\eta_{PQ} {\tilde V}'^P (\partial_- {\tilde Z}^Q +
{\tilde U}^Q)
={\tilde v'}_p \partial_- {\tilde X}^p + 
{\tilde v'}.{\tilde u}=0\;.
\eea
These four conditions imply that the number of the
degrees of freedom of the actions
(2) and (6) are equal. Since the vectors 
$v$ and ${\tilde v}$ are perpendicular
to the vectors $v'$ and ${\tilde v'}$, respectively,
these equations also can be obtained from the equations (18).
\subsection{The conformal ghosts}
The action of the conformal ghosts is
\bea
S_{cg} = \frac{1}{2\pi \alpha'}\int d^2\sigma (b \partial_+c
+{\tilde b}\partial_-{\tilde c})\;.
\eea
We request this action to be 
\bea
S_{cg} = \frac{i}{2\pi \alpha'}\int d^2\sigma \bigg{(}\eta_{PQ}
(\Psi^P_1 \partial_+\Theta^Q_1 + \Psi^P_2 \partial_-\Theta^Q_2)
\bigg{)}\;.
\eea
This action has all symmetries of the action (20). 
This is due to the fact that, 
it has two distinct copies of (20).

The Grassmannian variables $\{\Psi^P_{1,2} , \Theta^P_{1,2}\}$ have the 
following expressions in terms
of the worldsheet fermions $\{\psi^p , \theta^p\}$,
\bea
&~&\Psi^P_1 = e^P_{\;\;\;p} \psi^p_-\;,
\nonumber\\
&~&\Psi^P_2 = e^P_{\;\;\;p} \psi^p_+\;,
\nonumber\\
&~&\Theta^P_1 = e^P_{\;\;\;p} \theta^p_-\;,
\nonumber\\
&~&\Theta^P_2 = e^P_{\;\;\;p} \theta^p_+\;.
\eea
Therefore, the auxiliary action (21) can be written in terms of the
spinor fields
\bea
S_{cg}=\frac{i}{4\pi \alpha'}\int d^2\sigma
(G_{pq}{\bar \psi}^p \rho^a \partial_a \theta^q)\;,
\eea
where
$\psi^p=\left(\begin{array}{c}
\psi^p_-\\
\psi^p_+
\end{array} \right)$
and
$\theta^p=\left(\begin{array}{c}
\theta^p_-\\
\theta^p_+
\end{array} \right)$
are Majorana spinors.
We assumed that the vielbeins are independent of the fields
$\{X^p,{\tilde X}^p\}$, and hence they do not depend on the worldsheet
coordinates $\tau$ and $\sigma$. That is, the metric
$G_{pq}$ is constant. For matching the degrees of freedom of the actions
(20), (21) and (23) see the conditions (30).

Equality of the actions (20) and (21) gives
\bea
&~&b\partial_+c = i\eta_{PQ}\Psi^P_1 \partial_+ \Theta^Q_1 \;,
\nonumber\\
&~&{\tilde b}\partial_-{\tilde c} = i\eta_{PQ}\Psi^P_2 \partial_-
\Theta^Q_2 \;.
\eea
This is due to the common feature of these actions.
Consider the unit vectors $W^P$ and ${\widetilde W}^P$,
\bea
\eta_{PQ}W^PW^Q= \eta_{PQ}{\widetilde W}^P{\widetilde W}^Q =1\;.
\eea
Furthermore, define the Grassmann valued vectors
$\Lambda^P$ and ${\tilde \Lambda}^P$ with the following properties
\bea
\eta_{PQ} \Psi^P_1 \Lambda^Q = \eta_{PQ} \Psi^P_2 
{\tilde \Lambda}^Q = 0, 
\eea
\bea
\eta_{PQ} W^P \Lambda^Q \neq 0\;\;\;,\;\;\;
\eta_{PQ} {\widetilde W}^P {\tilde \Lambda}^Q \neq 0. 
\eea
Thus, for example, $\Lambda^P$ is perpendicular to
$\Psi^P_1$ but it is not perpendicular to $W^P$.

Insert the unit vectors in the left-hand sides of the equations (24).
Similar to the relations (10), (11) and (14), we obtain the mappings
\bea
&~&b=i\eta_{PQ}W^P \Psi^Q_1 =iw_p \psi^p_-\;,
\nonumber\\
&~&\partial_+c=\eta_{PQ}W^P (\partial_+\Theta^Q_1 + \Lambda^Q)
=w_p\partial_+\theta^p_- + w. \lambda\;,
\nonumber\\
&~&{\tilde b}=i\eta_{PQ}{\widetilde W}^P \Psi^Q_2
= i{\tilde w}_p \psi^p_+\;,
\nonumber\\
&~&\partial_-{\tilde c}=\eta_{PQ}{\widetilde W}^P (\partial_-  
\Theta^Q_2 +{\tilde \Lambda}^Q)
={\tilde w}_p\partial_-\theta^p_+ +{\tilde w}.{\tilde \lambda}\;.
\eea
The unit vectors $w_p$ and ${\tilde w}_p$ and the 
Grassmannian vectors 
$\lambda_p$ and ${\tilde \lambda}_p$, similar to the equation (15),
have definitions in terms of $\{W^P\}$, $\{{\widetilde W}^P\}$,
$\{\Lambda^P\}$ and $\{{\widetilde \Lambda}^P\}$, respectively.
The equations (27) imply that the inner products $w.\lambda$
and ${\tilde w}.{\tilde \lambda}$ are nonzero. Therefore,
according to the equations (28) 
the conformal ghosts appear as components of
some spinor fields. Equivalently, the worldsheet fermions in terms of
the conformal ghosts are
\bea
&~&\psi^p_- = -iw^p b\;,
\nonumber\\
&~&\partial_+ \theta^p_- = w^p\partial_+c-\lambda^p\;,
\nonumber\\
&~&\psi^p_+ = -i{\tilde w}^p {\tilde b}\;,
\nonumber\\
&~&\partial_- \theta^p_+ = {\tilde w}^p\partial_-{\tilde c}
-{\tilde \lambda}^p\;.
\eea

Let the unit vectors $W'$ and ${\widetilde W}'$ be perpendicular
to the vectors $W$ and ${\widetilde W}$, respectively. Therefore,
we have analog of the equations (19), $i.e.$,
\bea
&~&\eta_{PQ}W'^P \Psi^Q_1 =w'_p \psi^p_-=0\;,
\nonumber\\
&~&\eta_{PQ}W'^P (\partial_+\Theta^Q_1 + \Lambda^Q)=
w'_p\partial_+\theta^p_- + w'.\lambda=0\;,
\nonumber\\
&~&\eta_{PQ}{\widetilde W}'^P \Psi^Q_2= {\tilde w'}_p \psi^p_+=0\;,
\nonumber\\
&~&\eta_{PQ}{\widetilde W}'^P (\partial_-
\Theta^Q_2 + {\tilde \Lambda}^Q)
={\tilde w'}_p\partial_-\theta^p_+ + {\tilde w'}.{\tilde \lambda}=0\;.
\eea
These conditions give the same number of the degrees of freedom for the actions
(20), (21) and (23). From the equations (29) also we can obtain these
conditions. That is, use the unit vectors $w'$ and ${\tilde w'}$,
which are perpendicular to $w$ and ${\tilde w}$, respectively.

The physical states, extracted from the action (1), satisfy some
conditions. For example, they are BRST-invariant. 
If we substitute the super-conformal ghosts from (14)
and the conformal ghosts from (28),
in the BRST charge and physical states, 
we obtain the equivalent BRST charge and
physical states. The equivalent states under the 
equivalent BRST charge
are invariant. This procedure also holds for other conditions on
the physical states.
\subsection{Quantum consistency}

For verifying the quantization, 
we need the explicit forms of the fields
$\gamma, {\tilde \gamma}, c,$ and ${\tilde c}$.
The equations (14) and (28) give them as in the following
\bea
&~& \gamma = v_pX^p + \gamma_0(\sigma^-)+v.u \sigma^+\;,
\nonumber\\
&~& {\tilde \gamma} = {\tilde v}_p{\tilde X}^p
+{\tilde \gamma}_0(\sigma^+) +{\tilde v}.{\tilde u}\sigma^-\;,
\nonumber\\
&~& c = w_p \theta^p_- + c_0(\sigma^-) +w.\lambda \sigma^+\;,
\nonumber\\
&~& {\tilde c} = {\tilde w}_p \theta^p_+ + {\tilde c}_0(\sigma^+)
+{\tilde w}.{\tilde \lambda}\sigma^-\;,
\eea
where the vectors $\{v_p, {\tilde v}_p, w_p, {\tilde w}_p,
u_p, {\tilde u}_p, \lambda_p, {\tilde \lambda}_p\}$ are
considered independent of the coordinates 
$\sigma^-$ and $\sigma^+$.

The functions $\gamma_0$ and ${\tilde \gamma}_0$ commute 
with all fields and $c_0$
and ${\tilde c}_0$ anti-commute with all Grassmannian fields. 
Thus, the canonical quantization of the fields of the bosonic actions
(2) and (6) leads to the equations
\bea
[\gamma(\tau, \sigma), \beta(\tau , \sigma')]=
[{\tilde \gamma}(\tau, \sigma), {\tilde \beta}(\tau , \sigma')]
=4\pi i\alpha'\delta(\sigma-\sigma')\;,
\eea
\bea
[X^p(\tau, \sigma), \partial_\tau X^q(\tau , \sigma')]=
[{\tilde X}^p(\tau, \sigma), \partial_\tau {\tilde X}^q(\tau , \sigma')]
=4\pi i\alpha'G^{pq}\delta(\sigma-\sigma')\;.
\eea
Using the equations (14) and (31) and also 
the equations of motion of $\gamma$
and ${\tilde \gamma}$, the quantization (32) leads to 
the quantization (33) and vice-versa.

In the same way, the canonical quantization of the fields of the
actions (20) and (23) are 
\bea
\{c(\tau, \sigma), b(\tau , \sigma')\}=
\{{\tilde c}(\tau, \sigma), {\tilde b}(\tau , \sigma')\}=4\pi i\alpha'
\delta(\sigma-\sigma')\;,
\eea
\bea
\{\psi^p_-(\tau, \sigma), \theta^q_-(\tau , \sigma')\}=
\{\psi^p_+(\tau, \sigma), \theta^q_+(\tau , \sigma')\}=4\pi \alpha'
G^{pq}\delta(\sigma-\sigma')\;.
\eea
According to the equations (28) and (31), 
these quantizations are the same.
\section{Total superstring action}
Let express the worldsheet fields $X^p$ and ${\tilde X}^p$ in terms of
the new variables $x^p$ and ${\tilde x}^p$ as in the following
\bea
&~&X^p=A^p_{\;\;q}x^q + {\tilde B}^p_{\;\;q}{\tilde x}^q\;,
\nonumber\\
&~&{\tilde X}^p= B^p_{\;\;q}x^q +{\tilde A}^p_{\;\;q}{\tilde x}^q\;.
\eea
Apply these relations in the action (6) and 
then only keep the cross term of
the new variables. Therefore, the matrices 
$A$, ${\tilde A}$, $B$ and ${\tilde B}$
should satisfy the conditions
\bea
&~&A^T A +B^T B = 0\;,
\nonumber\\
&~&{\tilde A}^T {\tilde A} +{\tilde B}^T {\tilde B} = 0\;,
\nonumber\\
&~& A^T {\tilde B} +B^T {\tilde A} = {\bf 1}\;.
\eea
In other words, the actions (6) and (23) take the form
\bea
S_{g}=-\frac{1}{4\pi \alpha'}\int d^2\sigma\bigg{(}
G_{pq}(\eta^{ab} \partial_a x^p \partial_b {\tilde x}^q
-i{\bar \psi}^p \rho^a \partial_a \theta^q)\bigg{)}\;.
\eea
Obtaining the action (6) from the bosonic part of the action (38)
gives the relations
\bea
&~& A {\tilde A}^T +{\tilde B} B^T = 0\;,
\nonumber\\
&~& A {\tilde B}^T +{\tilde B} A^T = {\bf 1}\;,
\nonumber\\
&~& {\tilde A}B^T + B {\tilde A}^T = {\bf 1}\;.
\eea
These equations are not independent of the conditions (37).

However, the equations of motion, extracted from the action (38), are
\bea
\partial_a\partial^a x^p=\partial_a\partial^a {\tilde x}^p
= \rho^a\partial_a \psi^p= \rho^a\partial_a \theta^p=0\;.
\eea

To remove the reparametrization invariance of the action (38) we
can write it as combination of the matter form and the ghost form
\bea
S_{g} &=& -\frac{\mu}{4\pi \alpha'}\int d^2\sigma\bigg{(}
G_{pq}(\eta^{ab} \partial_a x^p \partial_b {\tilde x}^q
-i{\bar \psi}^p \rho^a \partial_a \theta^q)\bigg{)}
\nonumber\\
&~& +(1-\mu)\bigg{(}S_{scg}[Eq(2)]+S_{cg}[Eq(20)]\bigg{)}\;,
\eea
where $\mu$ is any real number. Each action in the second line does not
have the diffeomorphism invariance.

According to the action (38), the superstring
action (1) completely can be written in the matter form
\bea
S=-\frac{1}{4\pi \alpha'}\int d^2\sigma \bigg{(}
G_{{\bar \mu}{\bar \nu}}(\eta^{ab} \partial_a x^{\bar \mu}
\partial_b {\tilde x}^{\bar \nu}
-i{\bar \psi}^{\bar \mu} \rho^a \partial_a \theta^{\bar \nu} )\bigg{)}\;,
\eea
where the metric $G_{{\bar \mu}{\bar \nu}}$ is defined by
\bea
G_{{\bar \mu}{\bar \nu}}=
\left( \begin{array}{cc}
G_{\mu\nu} & 0\\
0 & G_{pq}
\end{array} \right)\;.
\eea
The coordinates $x^\mu$ and ${\tilde x}^\mu$ are defined by
$x^\mu ={\tilde x}^\mu = X^\mu$.
We also defined $\theta^\mu = \psi^\mu$.
The metrics $G_{\mu\nu}$ and $G_{pq}$ represent 9+1 and 1+1
signatures, respectively. We shall
see that the corresponding spacetime to the action (42) can have
the signatures 10+2 and 11+3 .
Now we study some symmetries of the action (38).
\section{Symmetries of the ghosts action in the matter form}
Since the symmetries of the matter part of 
the total superstring action (1) are
known, we concentrate on the symmetries of the ghost part, $i.e.$ $S_g$
in the form (38).
\subsection{The Poincar\'e-like symmetries}
Consider the global transformations
\bea
&~& \delta x^p = a^p_{\;\;q}x^q + b^p\;,
\nonumber\\
&~& \delta{\tilde x}^p={\tilde a}^p_{\;\;q}{\tilde x}^q+{\tilde b}^p\;,
\nonumber\\
&~& \delta \psi^p = a^p_{\;\;q}\psi^q\;,
\nonumber\\
&~& \delta \theta^p={\tilde a}^p_{\;\;q}\theta^q\;,
\eea
where $a_{pq}$ and ${\tilde a}_{pq}$ are antisymmetric
constant matrices. Under these transformations the action (38),
for ${\tilde a}_{pq}=a_{pq}$, is symmetric. The resulted current is
\bea
J^{pq}_a = \frac{1}{4\pi \alpha'}[
x^p \partial_a {\tilde x}^q- x^q \partial_a {\tilde x}^p
+{\tilde x}^p \partial_a x^q- {\tilde x}^q \partial_a x^p
+i({\bar \psi}^p \rho_a \theta^q-{\bar \psi}^q \rho_a \theta^p)]\;.
\eea
The equations of motion (40) imply that this is a conserved current
\bea
\partial^a J^{pq}_a = 0\;.
\eea
For the translations ${\tilde b}^p$ and $b^p$ the associated currents are
\bea
&~& P^p_a= \frac{1}{4\pi \alpha'} \partial_a x^p\;,
\nonumber\\
&~& {\tilde P}^p_a= \frac{1}{4\pi \alpha'} \partial_a {\tilde x}^p\;,
\eea
respectively. These currents also are conserved
\bea
\partial^a P^p_a = \partial^a {\tilde P}^p_a = 0\;.
\eea
For ${\tilde b}^p=b^p$ the action (38) again remains invariant and hence
we have the conserved current
\bea
{\cal{P}}^p_a= \frac{1}{4\pi \alpha'}(\partial_a x^p +
\partial_a {\tilde x}^p)\;.
\eea

Another symmetry of the action (38) is as follows
\bea
&~& \delta x^p = a^p_{\;\;q}{\tilde x}^q + b^p\;,
\nonumber\\
&~& \delta{\tilde x}^p={\tilde a}^p_{\;\;q} x^q+{\tilde b}^p\;,
\nonumber\\
&~& \delta \psi^p = a^p_{\;\;q}\theta^q\;,
\nonumber\\
&~& \delta \theta^p=-{\tilde a}^p_{\;\;q}\psi^q\;.
\eea
In these transformations the parameters $a_{pq}$ and ${\tilde a}_{pq}$
can be different. Therefore, there are two types of the group generators.
The associated currents are
\bea
&~& j^{pq}_a = \frac{1}{2\pi \alpha'}
(x^p \partial_a x^q- x^q \partial_a x^p
+i{\bar \psi}^p \rho_a \psi^q)\;,
\nonumber\\
&~& {\tilde j}^{pq}_a = \frac{1}{2\pi \alpha'}
({\tilde x}^p \partial_a {\tilde x}^q- {\tilde x}^q
\partial_a {\tilde x}^p
+i{\bar \theta}^p \rho_a \theta^q)\;.
\eea
Note that $j^{pq}_a$ corresponds to the parameter ${\tilde a}_{pq}$,
while ${\tilde j}^{pq}_a$ corresponds to $a_{pq}$.
These currents also satisfy the conservation laws
\bea
\partial^a j^{pq}_a = \partial^a {\tilde j}^{pq}_a = 0 \;.
\eea
The current equations for the 
translation parts are the same as (47)-(49).
Since the equations (44) and (50) are similar to the usual Poincar\'e
transformations, we call them as 
the {\it Poincar\'e-like} transformations.
\subsection{Worldsheet supersymmetries}

The matter part of the action (1) is symmetric under the worldsheet
supersymmetry transformations
\bea
&~& \delta X^\mu= {\bar \epsilon}\psi^\mu\;,
\nonumber\\
&~& \delta \psi^\mu =-i \rho^a \partial_a X^\mu \epsilon\;,
\eea
where $\epsilon$ is an infinitesimal constant spinor.
Now we study the various supersymmetries of the ghosts action (38).
\subsubsection{Bi-supersymmetries of the worldsheet}

Consider the following transformations 
\bea
&~&\delta x^p= {\bar \eta}\psi^p\;,
\nonumber\\
&~&\delta {\tilde x}^p= {\bar {\tilde \eta}}\theta^p\;,
\nonumber\\
&~& \delta \psi^p =-i \rho^a \partial_a x^p \eta\;,
\nonumber\\
&~& \delta \theta^p =-i \rho^a \partial_a {\tilde x}^p{\tilde \eta}\;,
\eea
where $\eta$ and ${\tilde \eta}$ are anticommuting constant
Majorana spinors. In fact, these are two independent transformations of
$(x^p , \psi^p ; \eta)$ and $({\tilde x}^p , \theta^p ; {\tilde \eta})$.
The associated supercurrents are
\bea
&~&J_a = \frac{1}{2}G_{pq}\rho^b\rho_a \psi^p \partial_b {\tilde x}^q\;,
\nonumber\\
&~&{\tilde J}_a=\frac{1}{2}G_{pq}\rho^b\rho_a \theta^p \partial_b x^q\;.
\eea
They correspond to the parameters $\eta$ and ${\tilde \eta}$,
respectively. Since there are two supercurrents,
we call this symmetry as
{\it bi-supersymmetry}. For ${\tilde \eta}= \eta$ the action (38) 
again remains symmetric. This leads to the current
\bea
{\cal{J}}_a = \frac{1}{2}G_{pq}\rho^b\rho_a (\psi^p \partial_b {\tilde x}^q
+\theta^p \partial_b x^q )\;.
\eea
The supercurrents (55) and (56) obey the conservation laws, $i.e.$,
\bea
\partial^a J_a = \partial^a {\tilde J}_a =\partial^a {\cal{J}}_a = 0 \;.
\eea

Besides the transformations (54), the action (38) also is
invariant under the following bi-supersymmetry transformations
\bea
&~&\delta x^p= {\bar \lambda}\theta^p\;,
\nonumber\\
&~&\delta {\tilde x}^p= {\bar {\tilde \lambda}} \psi^p\;,
\nonumber\\
&~& \delta \psi^p =-i \rho^a \partial_a {\tilde x}^p{\tilde \lambda}\;,
\nonumber\\
&~& \delta \theta^p =-i \rho^a \partial_a x^p \lambda\;,
\eea
where $\lambda$ and ${\tilde \lambda}$ are infinitesimal constant
real spinors. These transformations contain two independent parts
$(x^p , \theta^p ; \lambda)$ and $({\tilde x}^p , \psi^p ;{\tilde \lambda})$.
The associated supercurrents, corresponding to
the parameters $\lambda$ and ${\tilde \lambda}$, are
\bea
&~& k_a=\frac{1}{2}G_{pq}\rho^b\rho_a \theta^p \partial_b {\tilde x}^q\;,
\nonumber\\
&~& {\tilde k}_a=\frac{1}{2}G_{pq}\rho^b\rho_a \psi^p \partial_b x^q\;.
\eea
For ${\tilde \lambda}= \lambda$ we obtain analog of the current (56),
\bea
{\cal{K}}_a = \frac{1}{2}G_{pq}\rho^b\rho_a (\psi^p \partial_b x^q
+\theta^p \partial_b {\tilde x}^q )\;.
\eea
These supercurrents satisfy the conservation equations
\bea
\partial^a k_a = \partial^a {\tilde k}_a =\partial^a {\cal{K}}_a = 0 \;.
\eea
\subsubsection{The $N=2$ supersymmetry}

The $SO(10 , 2)$ covariant extension of the superstring is considered. For
this we change the worldsheet fermions $\{\psi^p ,\theta^p\}$ with
$\{\chi^p , {\tilde \chi}^p\}$ as in the following
\bea
&~& \psi^p=({\tilde B}^T)^p_{\;\;q}\chi^q + ({\tilde A}^T)^p_{\;\;q}
{\tilde \chi}^q\;,
\nonumber\\
&~& \theta^p=(A^T)^p_{\;\;q}\chi^q +(B^T)^p_{\;\;q}{\tilde \chi}^q \;.
\eea
The matrices in these equations are the same that appeared in the
equations (36), (37) and (39). 
Now we introduce these relations in the action (23).
Therefore, according to the action (6), the ghosts action takes the form
\bea
S_{g}=-\frac{1}{8\pi \alpha'}\int d^2\sigma \bigg{(}
G_{pq}[\eta^{ab}(\partial_a X^p \partial_b X^q
+\partial_a {\tilde X}^p \partial_b {\tilde X}^q)
-i({\bar \chi}^p \rho^a\partial_a \chi^q +
{\bar {\tilde \chi}}^p \rho^a\partial_a {\tilde \chi}^q)]\bigg{)}\;.
\eea
Obtaining the fermionic part, the conditions (37) and (39)
have been used.

This action manifestly describes the $N=2$ supersymmetry. 
The supersymmetry transformations are
\bea
&~&\delta X^p= {\bar \epsilon}\chi^p +
{\bar{\tilde \epsilon}}{\tilde \chi}^p\;,
\nonumber\\
&~&\delta {\tilde X}^p= {\bar \epsilon}{\tilde \chi}^p
-{\bar{\tilde \epsilon}}\chi^p\;,
\nonumber\\
&~& \delta \chi^p =-i \rho^a \partial_a X^p \epsilon
+i \rho^a \partial_a {\tilde X}^p {\tilde \epsilon}\;,
\nonumber\\
&~& \delta {\tilde \chi}^p =-i \rho^a \partial_a X^p{\tilde \epsilon}
-i \rho^a \partial_a {\tilde X}^p \epsilon\;.
\eea
The worldsheet fermions $\chi^p$ and ${\tilde \chi^p}$ form an $SO(2)$
doublet. 

In the appearance of $N=2$ supersymmetry the fields 
$\{X^p , \chi^p , {\tilde X}^p , {\tilde \chi}^p \}$, 
in the natural way appeared in the action (63). 
In other words, all fields in this action 
have originated from the conformal 
and super-conformal ghosts. 
That is, they have not been introduced by hand. Obtaining worldsheet 
supersymmetry with $N \geq 2$, 
some extra fields usually are added to action by hand.

In fact, the action (63) shows the other matter form of
the ghosts action. In this manner, the manifold ${\cal{M}}_g$
represents two-dimensional spacetime with the coordinates $\{X^p\}$
or $\{{\tilde X}^p\}$ and the signature 1+1.
This implies that the total target space for the superstring is 10+2
dimensional spacetime with the coordinates $\{X^\mu , X^p\}$ or
$\{X^\mu , {\tilde X}^p\}$.
As it was explained, this signature has origin in the conformal symmetry.
On the other hand, this space is
product of the ten-dimensional spacetime and the two-dimensional
manifold ${\cal{M}}_g$. Note that one direction of development of
supersymmetric theories is consideration of 12-dimensional theories
with the signature 10+2 \cite{2}.

The inverse of the fermions redefinition (62) is
\bea
&~& \chi^p=A^p_{\;\;q}\psi^q + {\tilde B}^p_{\;\;q}\theta^q\;,
\nonumber\\
&~& {\tilde \chi}^p=B^p_{\;\;q}\psi^q + {\tilde A}^p_{\;\;q}\theta^q\;.
\eea
Action of the operators $\partial_{\pm}$ on the equations (36) and
then comparing the resulted equations
with (65) give the following pairs of the super-partners
\bea
&~& X^p \leftrightarrow \chi^p \;,
\nonumber\\
&~& {\tilde X}^p \leftrightarrow {\tilde \chi}^p \;,
\nonumber\\
&~& x^p \leftrightarrow \psi^p \;,
\nonumber\\
&~& {\tilde x}^p \leftrightarrow \theta^p \;.
\eea
These are based on the actions (38) and (63) and their 
equations of motion.
That is, $\partial_+x^p$ ($\partial_-x^p$) acts in the same way as
$\psi^p_+$ ($\psi^p_-$), and so on.

Note that, application of the transformations (54) and (58) in the
equations (36) and (65) does not produce the transformations (64).
In other words, the transformations (54), (58) and (64) are independent.
\section{The signature 11+3 for the spacetime}

Define the block diagonal metric ${\cal{G}}_{mn}$ as
\bea
{\cal{G}}_{mn}=
\left( \begin{array}{ccc}
G_{\mu\nu} & 0 & 0\\
0 & \frac{1}{2}G_{pq} & 0\\
0 & 0 & \frac{1}{2}G_{p'q'} \\
\end{array} \right)\;.
\eea
In fact, the metrics $G_{pq}$ and $G_{p'q'}$ are equal. They have been
given by the equation (5). Therefore, the action (63) and the matter
part of the action (1) can be combined as in the following
\bea
S=-\frac{1}{4\pi \alpha'}\int d^2\sigma \bigg{(}
{\cal{G}}_{mn}(\eta^{ab} \partial_a Z^m \partial_b Z^n
-i{\bar \Omega}^m \rho^a \partial_a \Omega^n )\bigg{)}\;,
\eea
where the generalized coordinates $\{Z^m\}$ and the extended
worldsheet spinors $\{\Omega^m\}$ are defined by
\bea
&~& Z^\mu = X^\mu\;, \;\;\;Z^p = X^p\;, \;\;\;Z^{p'} ={\tilde X}^{p'}\;,
\nonumber\\
&~& \Omega^\mu = \psi^\mu\;, \;\;\;\Omega^p = \chi^p\;, \;\;\;
\Omega^{p'} ={\tilde \chi}^{p'}\;.
\eea
It is easy to see that this action has the $N=1$ supersymmetry.
The related transformations are analog of (53), $i.e.$,
\bea
&~& \delta Z^m= {\bar \varepsilon}\Omega^m\;,
\nonumber\\
&~& \delta \Omega^m =-i \rho^a \partial_a Z^m \varepsilon\;,
\eea
where the spinor $\varepsilon$ is real and constant.

As we explained, the metrics $G_{pq}$ and $G_{p'q'}$ are equal. 
For each of the metrics $G_{pq}$ and $G_{p'q'}$
the symmetry $SO(1,1)$ is added to the global Poincar\'e symmetry.
In other words, the conformal symmetry implies 
that the block diagonal metric
${\cal{G}}_{mn}$ and the
coordinates $\{Z^m\}$ describe 14-dimensional spacetime, with the
signature 11+3. In fact, the conformal symmetry of the system introduces
the extra coordinates. 

The line element of the spacetime associated to the action (68) is
\bea
ds^2 = {\cal{G}}_{mn}dZ^mdZ^n= G_{\mu\nu}dX^\mu dX^\nu+\frac{1}{2}
G_{pq}(dX^p dX^q + d{\tilde X}^p d{\tilde X}^q)\;.
\eea
On the other hand, this spacetime is product of the two manifolds
$M \times {\cal{M}}_g$, where $M$ is the 9+1 dimensional spacetime 
and ${\cal{M}}_g$ is 2+2 dimensional ghosts manifold.

Note that the $N=1$ supersymmetry in 11+3 
dimensions from the various point of 
view has been studied. For example, see the Ref.\cite{6}.
\section{Conclusions}
The super-conformal and conformal ghosts action, $i.e.$ $S_g$, was
expressed in the covariant matter form. The ghost fields also
were expressed in terms of the matter fields.
In other words, the conformal ghosts are equivalent to some spinor fields
and the super-conformal ghosts also can be represented by some  
scalar fields. 
We showed that the quantization of the action 
$S_g$ in the matter form and in the
ghost form are consistent, as expected.

We observed that the
bosonic fields of the matter form of the 
action can be interpreted as additional
coordinates of the spacetime.
The manifold of these extra dimensions is 1+1 or 2+2 
dimensional spacetime. 
Therefore, the total superstring action
corresponds to a twelve or fourteen-dimensional spacetimes with two
and three time-directions, respectively.

We studied some symmetries of $S_g$ in the matter form. Under two different
transformations, which are similar to the Poincar\'e transformations,
the action is invariant. 
Each of these symmetries gives the various conserved currents. 
Furthermore, for this action there are two
different bi-supersymmetries. A bi-supersymmetry contains
two parameters for the transformations and
hence two conserved supercurrents. 
In addition, we observed that the theory has
the $N=2$ supersymmetry. Finally we obtained the total superstring action,
with the $N=1$ supersymmetry, in the 11+3 dimensional spacetime.


\end{document}